\documentclass[prb,aps,twocolumn,showpacs,nobibnotes,epsf]{revtex4}
\usepackage{graphicx}% Include figure files
\usepackage{dcolumn}% Align table columns on decimal point
\usepackage{bm}% bold math
\usepackage{SIunits}
\usepackage{tabularx}

\begin{document}
\title{Transport Properties and Electronic Phase Diagram of Single-Crystalline Ca$_{10}$(Pt$_3$As$_8$)
((Fe$_{1-x}$Pt$_x$)$_2$As$_2$)$_5$}
\author{Z. J. Xiang, X. G. Luo, J. J. Ying, X. F. Wang, Y. J. Yan, A. F. Wang, P. Cheng, G. J. Ye, and X. H. Chen}
\altaffiliation{Corresponding author} \email{chenxh@ustc.edu.cn}
\affiliation{Hefei National Laboratory for Physical Science at Microscale and Department of
Physics, University of Science and Technology of China, Hefei, Anhui 230026, People's Republic
of China\\}

\begin{abstract}
Sizable single-crystalline samples of Ca$_{10}$(Pt$_3$As$_8$)((Fe$_{1-x}$Pt$_x$)$_2$As$_2$)$_5$ (the
10-3-8 phase) with 0$\leq x<$0.1 have been grown and systematically characterized via X-Ray diffraction,
magnetic, and transport measurements. The undoped sample is a heavily doped semiconductor with no sign of
magnetic order down to 2 K. With increasing Pt content, the metallic behavior appears and superconductivity
is realized for $x\geq$0.023. $T_{\rm c}$ rises to its maximum of approximately 13.6 K at the doping level
of $x\sim$0.06, and then decreases for higher $x$ values. The electronic phase diagram of the 10-3-8 phase was mapped out based on the transport measurements. The mass anisotropy parameter $\Gamma\sim$10 obtained from resistive measurements in magnetic fields
indicates a relatively large anisotropy in the iron-based superconductor family. This strong 2D
character may lead to the absence of magnetic order. A linear $T$-dependence of susceptibility at
high temperature is observed, indicating that spin fluctuations exist in the underdoped region as
in most of the Fe-pnictide superconductors.
\end{abstract}


\vskip 300 pt

\maketitle

\section{Introduction}
In the discovered high-temperature superconductors, which include cuprates and iron-based superconductors, superconductivity is often found in proximity to a magnetically ordered state. The parent compounds of cuprates
are antiferromagnetic insulators while for iron-based superconductors the parent compounds are antiferromagnetic "semimetals". By charge injection via chemical substitution, magnetic order is suppressed and superconductivity appears
in both of these two families. It is widely accepted nowadays
that there is an intimate association between magnetism and superconductivity in the high-temperature
superconductors. Different from the cuprates, in which Mott physics is dominant and the magnetic
order is a Heisenberg AFM order of localized spins, the magnetic order of the Fe-pnictides is spin-density wave (SDW) type and exhibits a significant itinerant character. Most of Fe-pnictide superconductors
have SDW region next to or overlapping with the superconducting region in their electronic phase diagrams
\cite{Chen, Luetkens, Zhao, Drew, Wang2}. Many theories suggest that the spin dynamics play a crucial role
in the pairing mechanism for the superconductivity in the Fe-pnictide superconductors\cite{Lee, Lumsden}, and it is conjectured
that AFM spin fluctuations mediate the s$_{\pm}$ pairing and are responsible for the high $T_{\rm c}$ in
Fe-pnictides\cite{Mazin, Eremin, Tesanovic}. However, there are several kinds of Fe-pnictide
superconductors that have no report about the existence of long-range magnetic order, such as LiFeAs
\cite{Borisenko} and so-called "perov-FeAs" materials. The latter is a group of layered
materials in which FeAs layers were separated by perovskite-type layers. The chemical formula of
perov-FeAs is either ($Ae$$_{n+1}$$M$$_n$O$_{3n-1-y}$)(Fe$_2$$Pn$$_2$) or ($Ae$$_{n+2}$$M$$_n$O$_{3n-y}$)(Fe$_2$$Pn$$_2$),
with $Ae$ = Ca, Sr, Ba and $M$ = Mg, Al, Sc, Ti, V, Cr, Co, etc.\cite{Wen1, Ogino1, Ogino2, Shirage1, Shirage2}, among which the highest $T_{\rm c}$ ever reported is $\sim$47 K for (Ca$_4$(Mg, Ti)$_3$O$_y$)(Fe$_2$As$_2$)
\cite{Ogino3}. All of these nonmagnetic Fe-pnictide materials are intrinsic superconductors that show
superconductivity in the stoichiometric compound. Recently a new type of layered Fe-pnictide
superconductor, Ca$_{10}$(Pt$_n$As$_8$)((Fe$_{1-x}$Pt$_x$)$_2$As$_2$)$_5$(the 10-$n$-8 phase, $n$ = 3, 4)
was discovered\cite{Ni, Nohara, Johrendt}. These materials have complex crystal structure
with triclinic symmetry (space group $P-1$), in which Fe$_2$As$_2$ layers alternate with Pt$_n$As$_8$
layers forming a -Ca-(Pt$_n$As$_8$)-Ca-(Fe$_2$As$_2$)- stacking. We noticed that for the 10-3-8 phase
the stoichiometric compound Ca$_{10}$(Pt$_3$As$_8$)(Fe$_2$As$_2$)$_5$ was non-superconducting and show no visible magnetic transitions, while electron doping through partial replacing Fe by Pt in the
Fe$_2$As$_2$ layers induced superconductivity. These characters were special in the family of Fe-pnictide superconductors.
In this paper, we report the results of a systematic study of the transport and magnetic properties
of single-crystalline Ca$_{10}$(Pt$_3$As$_8$)((Fe$_{1-x}$Pt$_x$)$_2$As$_2$)$_5$ in different doping regions,
and present a corresponding electronic phase diagram. All the data indicated that there is no magnetic order in
this system. The undoped sample is a semiconductor instead of a SDW semimetal. Superconductivity
emerges upon 5d-transition metal Pt substituting on the Fe site, as in the case of Pt-doped 122 type
Fe-pnictide superconductors\cite{Saha, Kirshenbaum}. $T_{\rm c}$ reaches its maximum $\sim$13.6 K at the doping
level $x\sim$0.06, and further doping slowly suppresses $T_{\rm c}$. The overdoped samples gradually exhibit
a phase separation so that the SC region is not a perfect dome-shape one. We also mentioned that the
AFM spin fluctuations still exist in this system as well as in other Fe-pnictide superconductors,
and the reason for the absence of AFM order might be ascribed to the highly anisotropy in the 10-3-8 phase.

\section{Experimental results}
\begin{figure*}[hp]
\centering
\includegraphics[width=0.95\textwidth]{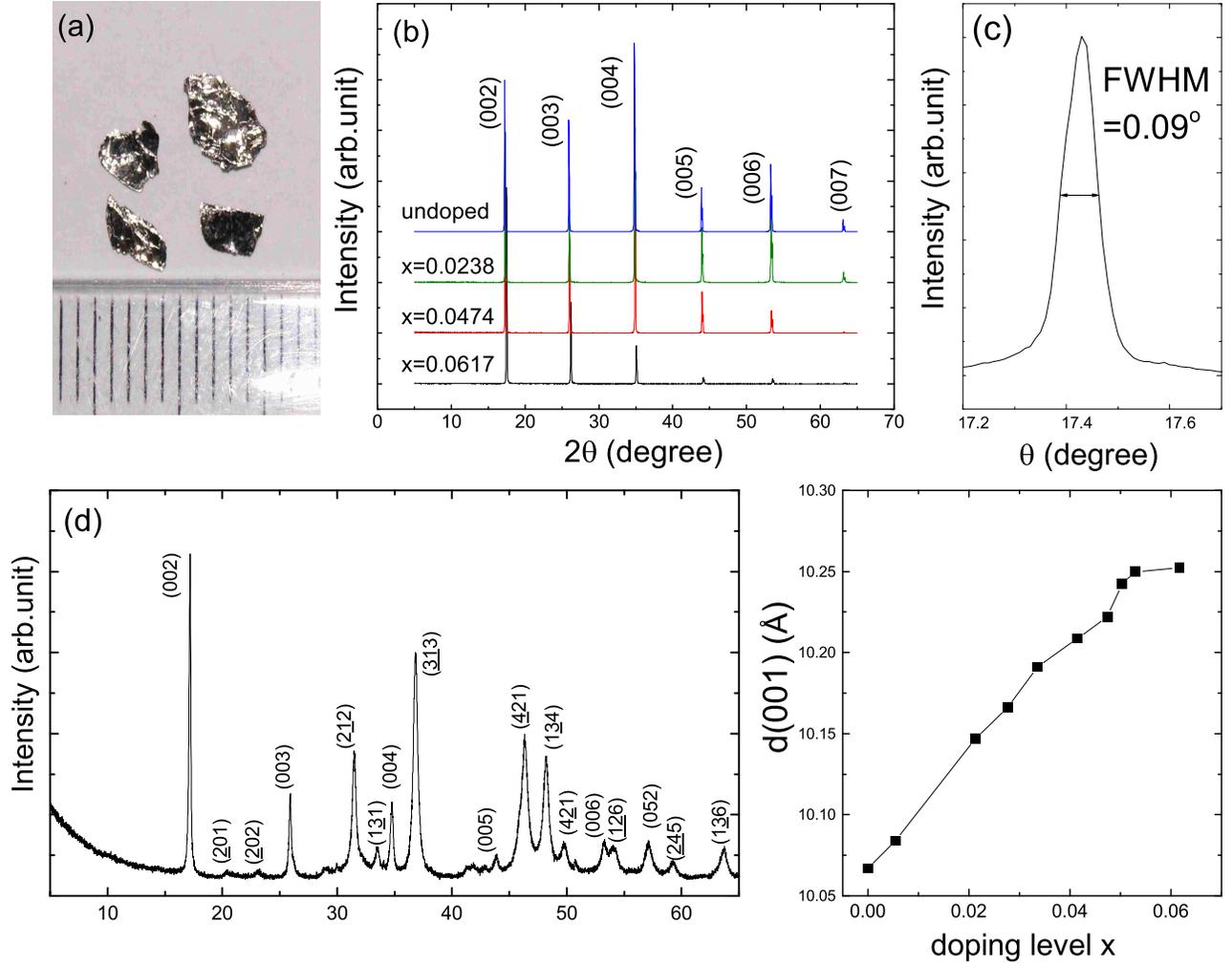}
\caption{(color online). (a) Photograph of single crystal pieces of Ca$_{10}$(Pt$_3$As$_8$)((Fe$_{1-x}$Pt$_x$)$_2$As$_2$)$_5$.
(b) X-ray diffraction pattern of single crystals. (c) Rocking curve of the (004) reflection. (d) X-ray powder
diffraction pattern of underdoped sample with doping level x$\approx$0.025. (e) Doping dependence of the interlayer spacing
of the Fe plane.} \label{fig1}
\end{figure*}

Single crystals of 10-3-8 sample were grown by self-flux method. Precursors
CaAs and FeAs were prepared by reacting the mixture of the element in the evacuated quartz
tubes at 923 K for 24 h, and at 973 K for 12 h, respectively. The starting material
CaAs, FeAs and Pt were mixed by a ratio of 2:2:(0.6+$x$)($x$ = 0, 0.1, 0.2, 0.3, 0.4)
in an Argon-filled glovebox, thoroughly ground and put into alumina crucibles. The crucibles were sealed in an evacuated quartz tube, and then heated to a temperature above 1273 K at a rate of 100 K/h.
For the superconducting samples, the best reaction temperature is 1323 K. The tubes were kept
at this temperature for 75 hours and then cooled to 1173 K at a rate of 4.5 K/h. Finally
the quartz tube was cooled in the furnace after shutting off the power. For the undoped
compound, we chose higher reaction temperature as 1423 K and the reaction time was prolonged
to 100 h, while the starting ratio of CaAs, FeAs and Pt was fixed at 2:2:1.05. After cooling
we obtained several dark gray granules with typical dimension 4$\times$4$\times$3 mm$^3$ together
with a little amount of gray powder in the crucibles. The shining plate-like 10-3-8 crystals
were cleaved from the internal of the granules. Single crystals were characterized by x-ray
diffraction (XRD) using Cu K$_\alpha$ radiation. The actual chemical composition of the single
crystals was determined by energy dispersive X-ray spectroscopy (EDS). The in-plane electrical
transport was measured with PPMS(Quantum Design Inc.) using the ac four-probe method. The Hall
effect was measured by the four-terminal ac technique. Magnetic susceptibility of superconducting
state was measured using a Quantum Design SQUID magnetometer. Normal state susceptibility was
measured by a vibrating sample magnetometer (VSM).

\section{Results and Discussions}
The typical size of cleaved single crystal is about 2$\times$3$\times$0.1 mm$^3$, as shown in
Figure 1(a). Figure 1(b) shows the the single crystal XRD patterns. Only
(00l) reflections are observed, indicating that the single crystals are in perfect (001) orientation.
The full width of half maximum (FWHM) in the rocking curve of the (004) peak is 0.09-0.20 degree, which indicates the
single crystals are of high quality. Determined by the results of EDS, the atomic ratios of single
crystals that cleaved from different granules in the same batch are slightly different,
but the chemical composition is approximately uniform within one granule. The EDS results
of samples from different starting material ratios are shown in table 1. The Pt doping
concentrations in the Fe$_2$As$_2$ layer were calculated from the relative atom ratio of iron and
platinum by assuming that there are no Pt vacancy or Fe substitution in the Pt$_3$As$_8$ layer,
according to the structure analysis in previous studies\cite{Ni, Nohara, Johrendt}.
When doping concentration x$>$0.075, it became rather difficult to get pure 10-3-8 phase. The XRD
pattern of single crystals in overdoped region usually shows two sets of (00l) diffraction
peaks, one corresponds to 10-3-8 phase and the other has larger (001) spacing (more than
10.3{\AA}),which can be attributed to the existence of so-called "10-4-8" phase\cite{Ni,Johrendt}
intergrowing with 10-3-8 phase. The sample No.12 in table 1 with $x$ = 0.0981, which is discussed
in this paper as overdoped sample, does not show the intergrowing phenomenon. We did not
succeed in growing single phase sample with doping level $x>$0.1.

\begin{table*}
\caption{Atom ratios in the 10-3-8 phase}
\def\temptablewidth{\textwidth}
{\rule{\temptablewidth}{1pt}}
\begin{tabular*}{\temptablewidth}{@{\extracolsep{\fill}}cccc} \hline
  Sample number & Starting ratio  & EDS results  & doping level x \\
   & Ca:Fe:Pt:As& Ca:Fe:Pt:As & \\ \hline
  1  & 2 : 2 : 1.05 : 4 & 2 : 2.083 : 0.598 : 3.558 & undoped \\ \hline
  2  & 2 : 2 : 0.6 : 4  & 2 : 1.958 : 0.643 : 3.577 & 0.0213\\
  3  &                  & 2 : 1.960 : 0.675 : 3.527 & 0.0238\\
  4  &                  & 2 : 1.932 : 0.667 : 3.510 & 0.0335\\ \hline
  5  & 2 : 2 : 0.7 : 4  & 2 : 1.946 : 0.682 : 3.540 & 0.0375\\
  6  &                  & 2 : 1.926 : 0.686 : 3.548 & 0.0415\\
  7  &                  & 2 : 1.957 : 0.707 : 3.560 & 0.0449\\ \hline
  8  & 2 : 2 : 0.8 : 4  & 2 : 1.909 : 0.696 : 3.494 & 0.0474\\
  9  &                  & 2 : 1.942 : 0.716 : 3.499 & 0.0502\\ \hline
  10 & 2 : 2 : 0.9 : 4  & 2 : 1.908 : 0.711 : 3.540 & 0.0530\\
  11 &                  & 2 : 1.888 : 0.727 : 3.550 & 0.0617\\ \hline
  12 & 2 : 2 : 1.0 : 4  & 2 : 1.858 : 0.820 : 3.558 & 0.0981\\ \hline
\end{tabular*}
 {\rule{\temptablewidth}{1pt}}
\end{table*}

Figure 1(d) shows the powder XRD pattern of underdoped 10-3-8 phase. The powder was obtained
by grounding single crystal pieces, and the Miller indices were marked according to a
triclinic(P-1) unit cell symmetry. The lattice parameters determined by powder diffraction
were estimated to be $a$ = 8.7608 {\AA}, $b$ = 8.7551 {\AA}, $c$ = 10.6831 {\AA}; $\alpha$ = 94.6823$^{\circ}$,
$\beta$ = 104.2267$^{\circ}$, $\gamma$ = 89.9874$^{\circ}$. These values are generally in accordance
with former results\cite {Ni}. Figure 1(e) presents the evolution trend of the interlayer distance of two neighboring
Fe-Fe square plane with Pt doping into FeAs layers. The interlayer distance (d(001)) increasing rapidly in the
underdoped region, and the variation slows down as approaching optimal doping. As for $x>$0.05,
the value of d(001) almost unchanged with Pt content.

\begin{figure}[ht]
\centering
\includegraphics[width=0.48\textwidth]{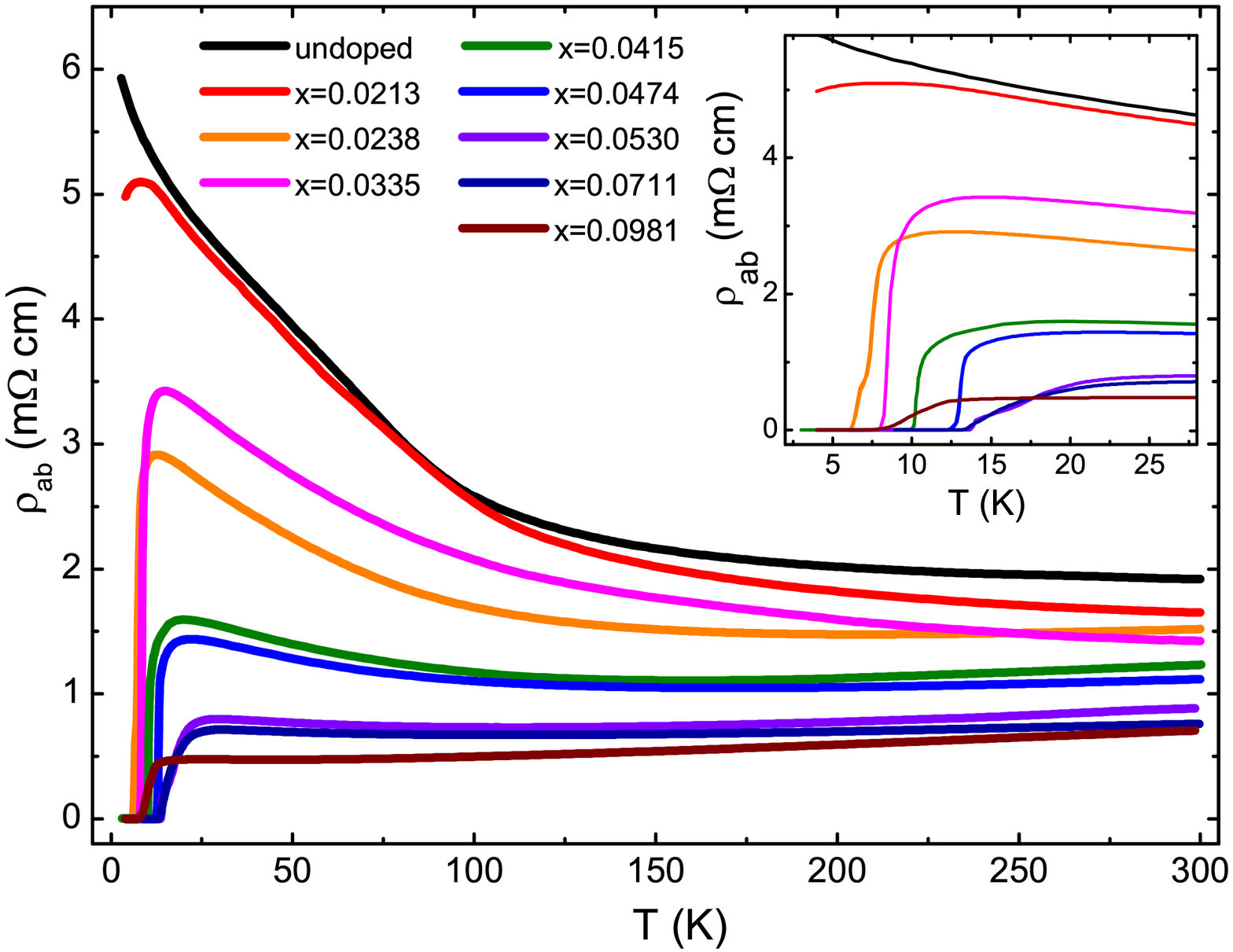}
\caption{(color online). The temperature dependence of resistivity for
Ca$_{10}$(Pt$_3$As$_8$)((Fe$_{1-x}$Pt$_x$)$_2$As$_2$)$_5$ samples.
The inset shows an expanded plot.} \label{fig2}
\end{figure}

Figure 2 shows the temperature dependence of in-plane resistivity of Ca$_{10}$(Pt$_3$As$_8$)
((Fe$_{1-x}$Pt$_x$)$_2$As$_2$)$_5$ single crystals. The behavior of resistivity of
the undoped sample Ca$_{10}$(Pt$_3$As$_8$)(Fe$_2$As$_2$)$_5$ is obviously different from those of the undoped
compound of the other iron-based superconductors. The parent compound of 1111 and 122 Fe-based
superconductors are so-called "semimetal" and an abnormal feature in resistivity is observed
at the SDW transition/structural transition\cite{Wang1, McGuire}. In this case, the undoped
sample is a heavily doped semiconductor, and the resistivity increases with cooling in the entire
temperature range from 300 K to 2 K. Below about 100 K a sharp increase in -d$\rho$/dT is observed,
but no magnetic anomaly was seen down to 2 K. The value of resistivity at room
temperature is of the same order of magnitude with polycrystalline LaFeAsO\cite{Kamihara},
one order of magnitude larger than BaFe$_2$As$_2$ single crystal, and one and two order of
magnitude smaller than the non-superconducting (Sr$_4$Sc$_2$O$_6$)(Fe$_2$As$_2$)\cite{Xie} and
the semiconducting phase of K$_x$Fe$_{2-y}$Se$_2$\cite{Yan}, respectively. With Pt doping into
the Fe$_2$As$_2$ layers, the resistivity gradually decreases and metallic behavior emerges at high temperature.
For samples with 0.015$<x<$0.023, the resistivity still shows a semiconducting behavior but decreases
below about 8 K without reaching zero, which can be regarded as a trace of superconductivity.
With further Pt doping, zero resistivity was observed. The zero resistivity temperature $T_{\rm c}$(0)
can reach the maximum of 13.6 K in the samples with $x$ = 0.0530 and $x$ = 0.0617, and decreases with further
doping when $x>$0.07. Since the overdoped region was affected by the coexistence of 10-3-8 and
10-4-8 phase, we could not obtain the overdoped sample of pure 10-3-8 phase in which superconductivity is fully
suppressed. In the whole superconducting region, most samples show a minimum in the normal-state
resistivity curve.  It should be mentioned that the temperature of resistivity minima, $T_{\rm min}$,
has a overall trend of shift to lower temperature upon Pt doping. The typical temperature
of resistivity minimum is about 150-200 K for the underdoped samples, 90-115 K for the optimally doped
samples and 50-70 K for the overdoped samples. Neither abrupt slope break at
$T_{\rm min}$ nor other anomalies which can be attribute to a phase transition have ever been observed
in the resistivity curves for all the samples, consistent with the previous reports\cite{Ni, Nohara}.
Below $T_{\rm min}$ the resistivity curves show an upturn, which becomes less pronounced upon Pt doping.
Similar phenomenon has been reported in 1111 Fe-based superconductors with element substitution
within FeAs layers\cite{Xu, Cao, Singh} and superconducting phosphides such as BaRh$_2$P$_2$\cite{Berry}.
This behavior has been explained as an effect of weak localization or spin-flip scattering.
However, the upturn is suppressed by Pt doping, which is contradict to the prediction of Anderson
localization theory\cite{Cao}, meanwhile, it is hard to accept the Kondo-like scenario since Pt
substitution does not introduce local moments as Co or Ni doping. Up to now the reason of
this upturn remains unclear.

\begin{figure}[ht]
\centering
\includegraphics[width=0.48\textwidth]{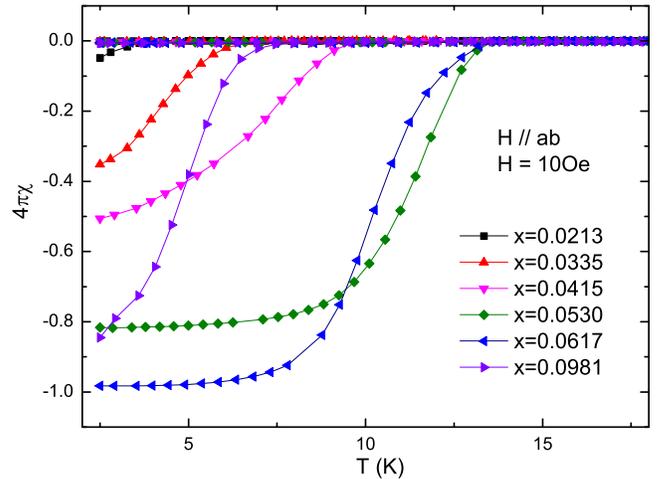}
\caption{(color online). The temperature dependence of in-plane magnetic susceptibility for superconducting
Ca$_{10}$(Pt$_3$As$_8$)((Fe$_{1-x}$Pt$_x$)$_2$As$_2$)$_5$ samples with magnetic field 10 Oe.} \label{fig3}
\end{figure}

Figure 3 presents the temperature dependence of magnetic susceptibility $\chi$ of the superconducting
samples measured under zero-field-cooling (ZFC) and field-cooling (FC) procedures by applying a
magnetic field of 10 Oe along the $ab$-plane at low temperatures. All the samples with $x<$0.02 show no
diamagnetic signal above 2 K(not shown). For sample with $x$ = 0.0213, in which zero resistivity
was not observed, the diamagnetism signal can already be observed below the temperature T$_c$ = 4.3 K,
even though the magnetic shielding fraction is estimated to be less than 5\%. Taking the non-uniform Pt distribution
in sample into account, we suggest that the edge of superconducting region should be at a doping
level between $x$ = 0.020 and $x$ = 0.025. The shielding fraction at 2.5 K exceeds 30\% for samples with
$x>$0.03, reaches 80\% for samples $x>$0.05 and approximately 100\% for the optimally doped sample with $x$ = 0.0617,
indicating bulk superconductivity in these samples. The superconducting
transition temperature $T_{\rm c}$ in magnetic measurement is consistent with the zero resistivity
temperature in the electric measurement. Samples with doping level of $x$ = 0.0530 and $x$ = 0.0617 have the
maximum transition temperature in the 10-3-8 phase with $T_{\rm c}$ = 13.6 K, which is in consistent with
the result reported by Kakiya et al.\cite{Nohara}.

\begin{figure*}[ht]
\centering
\includegraphics[width=0.95\textwidth]{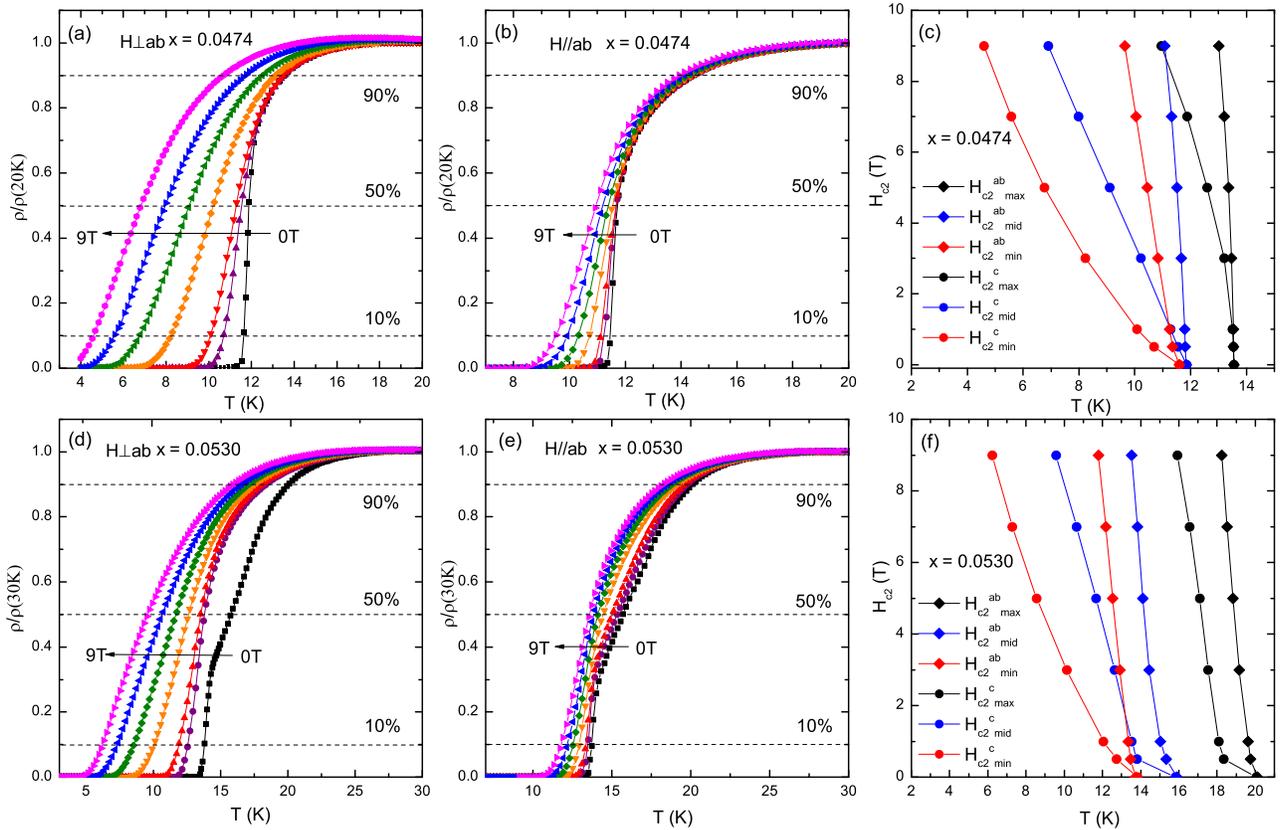}
\caption{(color online). In-plane electrical resistivity in magnetic fields (H = 0 T (black), 0.5 T (purple),
1 T (red), 3 T (orange), 5 T (olive), 7 T (blue) and 9 T (magenta), respectively) with (a)(d) H$\bot$ab and (b)(e)
H//ab and upper critical fields under different criteria for Ca$_{10}$(Pt$_3$As$_8$)((Fe$_{1-x}$Pt$_x$)$_2$As$_2$)$_5$
samples with (a)(b)(c) x = 0.0474 and (d)(e)(f) x = 0.0530.} \label{fig4}
\end{figure*}

In Fig. 4(a), (b), (d) and (e), we present the resistivity data of two superconducting samples in low-temperature
region under different fields. The samples studied were underdoped ones with $x$ = 0.0474 and $T_{\rm c}$ = 11.1 K,
and optimally doped sample with $x$ = 0.0617 and $T_{\rm c}$ = 13.6 K (the value of $T_{\rm c}$ was determined by
the susceptibility measurements). As there is a pronounced semiconductor-like behavior below $T_{\rm min}$
and preceding the onset of the superconducting transition, a round maximum is formed at low temperature.
The drop of resistivity below the temperature of this maximum is not very sharp. The interval between the maximum and the temperature
at which resistivity reaches zero is as wide as about 15 K even in the optimally doped sample (see
the inset of Fig. 2). As a result, it is rather difficult to determine the onset temperature
of superconductivity from resistivity measurement. We chose three criteria of $T_{\rm c}$ as 90\%, 50\% and 10\%
of the normal state resistivity (determined as the local resistivity maxima at low temperature), and
defined three critical field $H_{\rm C2}^{\rm max}$, $H_{\rm C2}^{\rm mid}$ and $H_{\rm C2}^{\rm min}$ following the three
criteria respectively. All the critical fields are shown in Fig. 4(c) and (f). For the underdoped
sample $x$ = 0.0474, the behavior of critical fields are sensitive to the used criterion. When magnetic field
applied perpendicular to $ab$ plane, the $H_{\rm C2}^{\rm max}$ shows a negative curvature while $H_{\rm C2}^{\rm min}$
shows obviously positive curvature, and $H_{\rm C2}^{\rm mid}$ has a nearly linear $T$-dependence. As for the
case of $H$//$ab$, the results are similar but the curvatures are not so obvious. By using Werthamer-Helfand-Hohenberg
formula\cite{Werthamer}, the upper critical field at zero-temperature can be estimated from
the initial slope (d$H_{\rm C2}$/d$T$)$_{T=T_{\rm c}}$. Under the 50\% criterion, the value of $H_{\rm C2}$(0) is about
143.4 T for the configuration of $H$//$ab$ and about 14.13 T for $H\bot ab$. The anisotropy parameter $\Gamma$ =
$H_{\rm C2}$$^{//}$/$H_{\rm C2}$$^{\bot}$ is derived to be about 10, which is much larger than those
of NdFeAsO$_{0.82}$F$_{0.18}$ ($\Gamma\leq$ 6)\cite{Jia} and doped Ba-122 superconductors ($\Gamma\sim$ 1.5-2 for
Ba(Fe$_{0.9}$Co$_{0.1}$)$_2$As$_2$\cite{Yamamoto} and Ba$_{0.6}$K$_{0.4}$Fe$_2$As$_2$\cite{Wen4}). Although
the application of WHH model is questionable since this material was proved to be a multi-band system\cite{Hasan},
the results at least indicated that the anisotropy of 10-3-8 phase is probably to be larger than those
in 1111- and 122-type Fe-pnictides. The negative curvature of $H_{\rm C2}^{\rm max}$ is not common in the iron-based
superconductors, and it may be affected seriously by the magnetoresistance of normal state since it is difficult
to fix the onset of superconductivity. On the other hand, $H_{\rm C2}^{\rm min}$ could be interpreted as the
irreversibility field, and the upward behavior resembles those in LaFeAsO$_{0.89}$F$_{0.11}$\cite{Hunte} and
Sr$_4$V$_2$O$_6$Fe$_2$As$_2$\cite{Wen1}. For the optimally doped sample, an upward curvature
was observed in all the critical field curves, and especially distinct for $H\bot ab$. Similar behavior
has been reported in cuprates\cite{Osofsky}, MgB$_2$\cite{Mueller} and Fe-based 1111 superconductors\cite{
Jia, Jaroszynski}. In 1111 Fe-pnictides this upward curvature was usually considered as a result of two-band
effect\cite{Hunte, Jaroszynski}. However, in the 10-3-8 phase the critical field curves show an upward bending
near $T_{\rm c}$, which is more pronounced than most of the other Fe-pnictide superconductors. This anomalous
upturn has been theoretically interpreted as an effect of the two-dimensional nature and being associated with
anisotropic Ginzburg-Landau behavior in the dirty limit\cite{Schneider}. Those properties of upper critical
fields in the underdoped and optimally doped samples have been confirmed by the measurement on other several
pieces of crystals with approximative doping concentration, and the shape of curves showed quite weak sample dependence.
Additionally, the large interval exists between the resistivity curves under zero-field and $H$ = 0.5 T in
optimally doped samples might be due to the inhomogeneity of Pt distribution, which formed small regions
with higher $T_{\rm c}$ and in which the superconductivity is easily suppressed by low magnetic field. Nonetheless, the
data of XRD and susceptibility afford no evidence that supports this assumption.

\begin{figure}[t]
\centering
\includegraphics[width=0.48\textwidth]{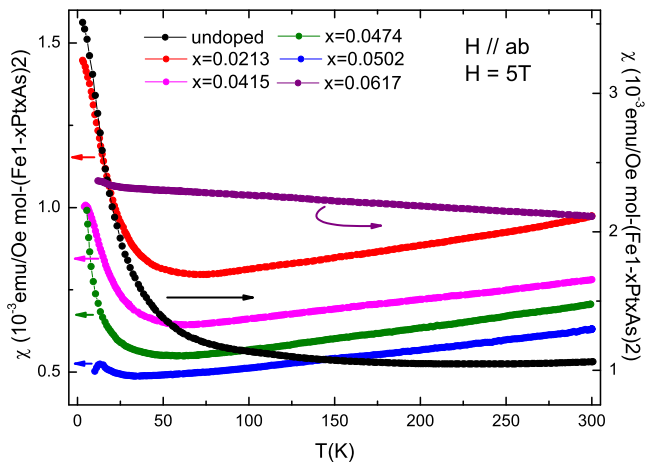}
\caption{(color online). The temperature dependence of in-plane magnetic susceptibility for
Ca$_{10}$(Pt$_3$As$_8$)((Fe$_{1-x}$Pt$_x$)$_2$As$_2$)$_5$ single crystals under $H$ = 5 T.} \label{fig5}
\end{figure}

In order to confirm that there is no magnetic transition existing in the 10-3-8 phase, the normal-state
susceptibility measurements were performed. Temperature dependence of the in-plane susceptibility (measured with magnetic field lying within the $ab$-plane) for samples with different Pt concentrations in the temperature range from 2 K to 300 K under $H$ = 5 T are plotted in Fig. 5.
For the undoped sample, the susceptibility $\chi$ first decreases
when cooling from 300 K, and below about 270 K the slope diminishes gradually and then changes its sign smoothly
at about 220-230 K. The temperature dependence of susceptibility is rather weak from 300 K to 100 K, and below
$\sim$75 K the susceptibility show a Curie-like upturn. Again, no magnetic anomalies could be observed in
the whole temperature region from 300 K to 2 K, which is different from iron-based parent compounds of 1111 and 122 systems. The paramagnetic behavior in the low-temperature region has been also reported in layered
non-superconducting Fe-oxypnictides Sr$_3$Sc$_2$O$_5$Fe$_2$As$_2$\cite{Wen2} and Sr$_4$Sc$_2$O$_6$Fe$_2$As$_2$
\cite{Xie}. For the underdoped superconducting samples, the in-plane susceptibility
exhibit a $T$-linear behavior above $\sim$70 K, which is similar to those observed at high temperature in the underdoped LaO$_{1-x}$F$_x$FeAs
and 122-materials $Ae$Fe$_2$As$_2$ ($Ae$ = Ca, Sr, Ba)\cite{Wang1, Wu, Wang2, Klingeler}. The slope of $T$-linear susceptibility
decreases slightly upon Pt doping (with 6.63$\times$10$^{-7}$ emu mol$^{-1}$ K$^{-1}$ for $x$ = 0.0213,
and 5.57$\times$10$^{-7}$ emu mol$^{-1}$ K$^{-1}$ for $x$ = 0.0502). The $T$-linear behavior has been explained
as an effect of strong ($\pi,\pi$) SDW fluctuations, and the slope is determined by the square of the SDW
amplitude with nesting momentum ${\bf Q}$ = ($\pi,\pi$)\cite{Korshunov}. Since the slope of $T$-linear susceptibility
in 10-3-8 phase lies between the value of the slope of Ni-doped LaFeAsO\cite{Cao} and Co-doped  BaFe$_2$As$_2$
\cite{Wang2}, we can conclude that even there is no SDW ordering in the 10-3-8 phase, strong
AFM spin fluctuations still exist in this system and may play an important role in inducing
superconductivity. At low temperature, the susceptibility of all the underdoped samples shows an
obvious upturn, which has also been reported in many other Fe-pnictide superconductors\cite{Xu, Cao, Maple}.
We attempted to fit the low temperature susceptibility of both undoped and underdoped samples using
Curie-Weiss formula. For underdoped samples we obtained small effective moments ($\sim$0.1 $\mu$$_{\rm B}$ per Fe site)
which shows considerable sample dependence, while for the undoped samples the Curie constant is one order of
magnitude larger. Thus we believe that the Curie-Weiss-type behavior in the underdoped samples is likely to be
extrinsic and could be ascribed to impurities and defects, as the case in 1111-materials, while in the
semiconducting undoped compound the paramagnetic behavior at low temperature might be intrinsic as in non-superconducting 32522 and
42622 systems. For the optimally doped and slightly overdoped samples, $T$-linear behavior is broken and
$\chi$ is nearly temperature independent down to the onset of superconductivity. Both the behavior and the
magnitude are similar to those in optimally doped 122 type Fe-pnictide superconductors\cite{Saha, Kumar}.

\begin{figure}[t]
\centering
\includegraphics[width=0.48\textwidth]{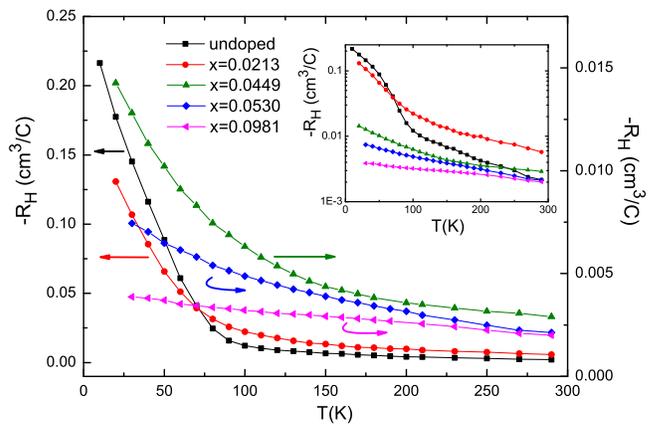}
\caption{(color online). The temperature dependence of Hall coefficient R$_H$ for
Ca$_{10}$(Pt$_3$As$_8$)((Fe$_{1-x}$Pt$_x$)$_2$As$_2$)$_5$ samples. The inset shows the same data on a
logarithmic scale.} \label{fig6}
\end{figure}

Figure 6 shows the temperature-dependent Hall coefficients for Ca$_{10}$(Pt$_3$As$_8$)((Fe$_{1-x}$Pt$_x$)$_2$As$_2$)$_5$
crystals with different Pt content. We checked the linearity in $H$ of the Hall voltage up 5 T. For all of the
samples the Hall coefficient $R_{\rm H}$ remains negative in whole temperature regime from $T_{\rm c}$ to 300 K, which
indicates that electron-type charge carriers dominate the conduction in all the samples. The absolute value of
$R_{\rm H}$ of the undoped sample is about twice as large as that of SmFeAsO\cite{Liu} at low temperature, but there
is no anomaly in the slope of the $R_{\rm H}$, which is related to magnetic transition. Nonetheless, the Hall coefficient of undoped sample
as well as the underdoped samples shows a strong temperature dependence at low temperatures, which suggests
either a strong multiband effect or a spin related scattering effect\cite{Wen2}. With increasing Pt doping level, this temperature
dependence becomes moderate and almost vanishes for the overdoped samples. The Hall concentration $n_{\rm H}$ = 1/($eR_{\rm H}$),
which represents carrier concentration in the single band model, however, does not follow a monotonic doping
dependence at high temperature. As shown in the inset of Fig. 6, underdoped samples with $x$ = 0.0213 and $x$ = 0.0449
have larger $R_{\rm H}$, that is, smaller Hall concentration than the undoped sample at room temperature. With enhancing Pt content, the
Hall concentration first decreases in underdoped region and then
turns to increase upon further doping. The turning point depends on temperature. All of these behaviors,
except for the absence of SDW transition, are similar to those in Ba(Fe$_{1-x}$Co$_x$)$_2$As$_2$
\cite{Wen3, Alloul}, which could be explained under a multi-band model as the competing effect of carrier
doping and hole mobility decreasing in the underdoped region. Besides, the Hall concentration for
the optimal doping at 200 K ($\sim$ 0.16$e$ per Fe) is almost the same as that in the Co-doped Ba-122 system, but is about
twice as large as that in the F doped SmFeAsO system.

\begin{figure}[b]
\centering
\includegraphics[width=0.48\textwidth]{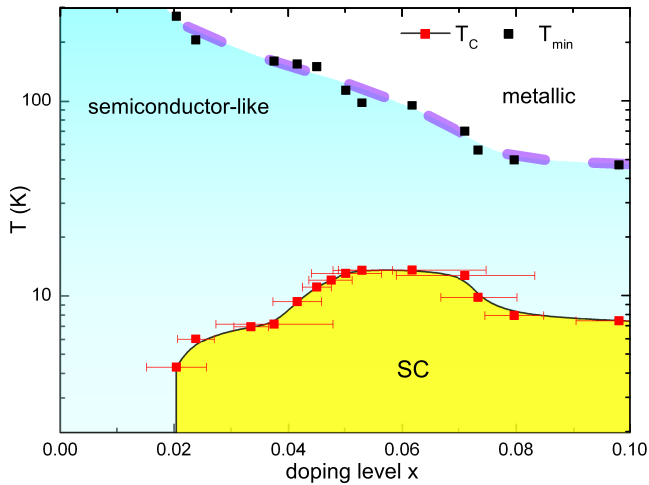}
\caption{(color online). Electronic phase diagram for Ca$_{10}$(Pt$_3$As$_8$)((Fe$_{1-x}$Pt$_x$)$_2$As$_2$)$_5$.
T$_{min}$ indicates the resistivity minimum temperature. T$_C$ was determined by susceptibility measurement.}
\label{fig7}
\end{figure}

Based on the data of transport and magnetic measurements shown above, an electronic phase diagram for
Ca$_{10}$(Pt$_3$As$_8$)((Fe$_{1-x}$Pt$_x$)$_2$As$_2$)$_5$ system was summarized in Fig. 7. The undoped compound,
Ca$_{10}$(Pt$_3$As$_8$)(Fe$_2$As$_2$)$_5$, is a heavily doped semiconductor without magnetic ordering,
different to the parent compounds of 1111 and 122 Fe-pnictide superconductors, which are SDW-type bad metal. The Pt
substitution on Fe site dopes electrons to the Fe$_2$As$_2$ layer, as is proved by the Hall coefficient
measurements. About 2\% Pt doping begins to introduce superconductivity, and $T_{\rm c}$ reaches its maximum $T_{\rm c}^{\rm max}$ = 13.6 K
in the doping range 0.050$<x<$0.065. Further Pt doping makes $T_{\rm c}$ decrease slowly. With the doping level up to $x\sim$0.1 superconducting transition can still be observed at about 7.4 K and further doping leads the single 10-3-8 phase
unable to be obtained. Therefore, the superconducting phase region is extremely asymmetric as in the Pt doped
Ba-122 system\cite{Zhu}. The normal state is divided by the line of $T_{\rm min}$ into semiconducting and
metallic regions, which is similar to Co doped and Ni doped 1111 systems\cite{Xu,Cao}. The most
extraordinary aspect of the phase diagram is the absence of SDW region which exists in the underdoped
side of the electronic phase diagram for all the 11, 1111 and 122 Fe-pnictide materials\cite{Canfield, Lumsden}.
In the former phase diagram established by Cho et al.\cite{Cho}, the magnetic and superconducting phases
are clearly separated. In this work we performed resistivity, susceptibility and Hall coefficient
measurements, all of the data indicate the fact that there is actually neither SDW nor other type of magnetic
order existing in the phase diagram of 10-3-8 system.

As mentioned above, static magnetic order is also absence in LiFeAs and "perov-FeAs" compounds. In
the latter case, due to the large thickness of perovskite-type blocking layer,
the distance between two nearest Fe$Pn$ layers is more than $\sim$13 {\rm \AA}
\cite{Kinouchi}, which is much larger than other types of layered iron-pnictides. It is believed that the
much stronger two-dimensional character compared to other Fe-pnictide superconductors, which causes
relatively weak magnetic coupling between FeAs layers, is destructive to the antiferromagnetic correlation
between the moments of Fe ions in the neighboring FeAs layers, and then prevents the system from forming a long
range magnetic order\cite{Wen2, Xie, Sefat, Munevar, Kinouchi}. In the 10-3-8 phase, the distance
between two neighboring FeAs layers is about 10.2 {\AA}, which is smaller than that in perov-FeAs materials
but still larger than in the 1111 materials (d$\sim$8.4-8.9 {\AA}) and 122 materials (d$\sim$5.8-6.6 {\AA}). The highly
anisotropic 2D nature of Ca$_{10}$(Pt$_3$As$_8$)((Fe$_{1-x}$Pt$_x$)$_2$As$_2$)$_5$ is already indicated by
our anisotropy parameter studies of upper critical field and the similar result by Ni et al\cite{Ni}. Thus it is possible that the weak
interlayer coupling suppresses the antiferromagnetic order in the 10-3-8 phase as in perov-FeAs materials.
However, the linear T-dependence of susceptibility at high temperature indicates that antiferromagnetic
spin fluctuation still exists in the 10-3-8 phase although magnetic ordering is suppressed. Strong
antiferromagnetic spin fluctuation has been observed in LiFeAs\cite{Taylor} and (Ca$_4$Al$_2$O$_{6-y}$)
(Fe$_2$As$_2$)\cite{Kinouchi}, both of which have no magnetic order close to or coexisting with
superconductivity. Therefore it is reasonable to conclude that there is also crucial relationship
between AFM spin fluctuations and superconductivity in these materials, including the 10-3-8 phase,
as in the other existing Fe-pnictide superconductors which have magnetic order in their parent compounds.

\section{Conclusion}
In summary, high quality Ca$_{10}$(Pt$_3$As$_8$)((Fe$_{1-x}$Pt$_x$)$_2$As$_2$)$_5$ single crystals with doping
level 0$\leq$x$<$0.1 were successfully grown by self-flux method. A systematic study of the transport
and magnetic propertis of the single crystal samples was performed, and an electronic phase diagram was
established. The undoped sample was a semiconductor without any type of magnetic order. Pt substitution
on the Fe site dopes electrons into the Fe$_2$As$_2$ layers and introduces of metallic resistivity behavior and
superconductivity. In the phase diagram there is no SDW region, which is a notable difference
to the phase diagram of 1111 and 122 type Fe-pnictide materials. We argued that the absence of long-range AFM order is due to the strong 2D character of this system that revealed by the relatively
large anisotropy parameter $\Gamma$. The extremely anisotropic nature weakens the interlayer
coupling, as in the perovskite-type layered Fe-based compounds. Apart from that, the properties of
superconducting samples were similar to other electron doped Fe-pnictide superconductors, indicating
that the properties of this system is dominated by Fe$_2$As$_2$ layers. This is in consistent with previous
results that the Pt$_3$As$_8$ layers couple only weakly with the Fe$_2$As$_2$ layers, and the contribution
of density of states at the Fermi level from Pt is rather small\cite{Hasan, Johrendt}. For underdoped
samples, the magnetic susceptibility shows T-linear dependence in a wide temperature range, indicating
strong magnetic fluctuation in this system. Thus it suggests that the mechanism of superconductivity of
Ca$_{10}$(Pt$_3$As$_8$)((Fe$_{1-x}$Pt$_x$)$_2$As$_2$)$_5$ is similar to that in other iron-based superconductors.
Being a special member of layered Fe-pnictide superconductor family with no magnetic order and shows variation
of ground state from paramagnetic semiconductor to superconductor controlled by electron doping,
the 10-3-8 phase is a good candidate to study the interplay between magnetic and superconductivity
in Fe-pnictide superconductors. Further research on the magnetic fluctuation in this system may
help to understanding the mechanism and nature of high-temperature superconductivity.

\section*{Acknowledgements}This work is supported by the National Natural Science Foundation of China
(Grant No. 11190020 and No. 51021091), National Basic Research Program of China (973 Program, Grant No. 2012CB922002 and No.
2011CB00101) and Chinese Academy of Sciences.

\end{document}